\documentclass[preprint,12pt]{elsarticle}
\journal{Nuclear Physics B}
\usepackage{graphicx}

\usepackage{lineno,hyperref}
\modulolinenumbers[5]
\usepackage{array}
\usepackage{microtype}
\usepackage{float}
\usepackage{amsfonts}
\usepackage{mathrsfs}
\usepackage{epstopdf}
\begin{document}
\begin{frontmatter}

\title{Study of gravastars under $f(\mathbb{T})$ gravity}

\author[addr1]{Amit Das}\ead{amdphy@gmail.com}
\author[addr2]{Shounak Ghosh}\ead{shounak.rs2015@physics.iiests.ac.in}
\author[addr2]{Debabrata Deb}\ead{ddeb.rs2016@physics.iiests.ac.in}
\author[addr3]{Farook Rahaman}\ead{rahaman@associates.iucaa.in}
\author[addr1]{Saibal Ray}\ead{saibal@associates.iucaa.in}

\address[addr1]{Department of Physics, Government College of Engineering and Ceramic Technology, Kolkata 700010, West Bengal, India}
\address[addr2]{Department of Physics, Indian Institute of Engineering Science and Technology, B. Garden, Howrah 711103, West Bengal, India}
\address[addr3]{Department of Mathematics, Jadavpur University, Kolkata 700032, West Bengal, India}

\date{Received: date / Accepted: date}

\begin{abstract}
In the present paper we propose a stellar model under the 
$f(\mathbb{T})$ gravity following the conjecture of Mazur-Mottola~[Report number:
LA-UR-01-5067 (2001); Proc. Natl. Acad. Sci. USA 101 (2004) 9545]
known in literature as {\it gravastar}, a viable alternative to the black hole. 
This gravastar has three different regions, viz., (A) Interior core region, (B)
Intermediate thin shell, and (C) Exterior spherical region. 
It is assumed that in the interior region the fluid pressure is equal to 
a negative matter-energy density providing a constant repulsive force over the
spherical thin shell. This shell at the intermediate region 
is assumed to be formed by a fluid of ultrarelativistic plasma 
and the pressure, which is directly proportional to the matter-energy 
density according to Zel'dovich's conjecture of stiff fluid~[Zel’dovich, Mon. Not.
R. Astron. Soc. 160 (1972) 1], does nullify the repulsive force exerted by 
the interior core region for a stable configuration. On the other hand, the 
exterior spherical region can be described by the exterior 
Schwarzschild-de Sitter solution. With all these specifications we have found 
out a set of exact and singularity-free solutions of the gravastar 
which presents several physically interesting as well as valid features 
within the framework of alternative gravity.
\end{abstract}

\end{frontmatter}

\section{Introduction}
Mazur and Mottola~\cite{Mazur2001,Mazur2004} proposed an interesting stellar model which 
they termed as {\it gra}vitationally {\it va}cuum {\it star} (gravastar) as an alternative 
to end point of a star due to gravitational collapse, i.e., the black hole. They constructed the 
model of the gavastar as a cold, dark and compact object of the interior de Sitter condensate
phase by extending the idea of Bose-Einstein condensation.

According to Mazur and Mottola, the gravastar has specifically three different regions. The interior of the
gravastar is surrounded by a thin shell of ultrarelativistic matter whereas the exterior is completely
vacuum which can be described by Schwarzschild geometry perfectly. So, the entire system of gravastar
can be defined by different equation of state (EOS) as follows: \\

(A) Interior ($0 \leq r < r_1 $):~$ p = -\rho $,

(B) Shell ($ r_1 \leq r \leq r_2 $):~$ p = +\rho $,

(C) Exterior ($ r_2 < r $ ):~$ p = \rho =0. $ \\ 

Here, the  shell is assumed to be very thin lying within the range $ r_1 \leq r \leq r_2 $ where $r_1\equiv~D$ and $r_2\equiv~D+\epsilon$ are the interior and exterior radii of the gravastar respectively, $\epsilon$ ($\ll1$) being the thickness of the shell.

There are lots of works related to the gravastar available in literature based on several mathematical as
well as physical issues. But most of these works are done in the framework of Einstein's general relativity~\cite{Mazur2001,Mazur2004,Visser2004,Cattoen2005,Carter2005,Bilic2006,Lobo2006,DeBenedictis2006,Lobo2007,Horvat2007,Cecilia2007,Rocha2008,Horvat2008,Nandi2009,Turimov2009,Usmani2011,Lobo2013,Bhar2014,Rahaman2015}.
Though Einstein's general relativity is one of the cornerstones of modern theoretical physics and has always proved
to be fruitful for uncovering so many hidden mysteries of the Nature, yet it presents some shortcomings in theoretical as
well as observational and experimental viewpoints. Observational evidences of the expanding universe with acceleration as well as the existence of dark matter has posed a theoretical challenge to this theory~\cite{Ri1998,Perl1999,Bern2000,Hanany2000,Peebles2003,Paddy2003,Clifton2012}.
Hence, several alternative gravity theories such as $f(R)$ gravity, $f(\mathbb{T})$ gravity, $f(R,\mathcal{T})$ gravity etc.
have been proposed time to time. All these theories can be considered necessary to describe the structure formation as well as evolution of the stellar systems of the Universe. As interesting applications of these theories one may look into the works~\cite{Yousaf2016a,Yousaf2018,Yousaf2019a} for the study of gravitational collapse of the gravitating system as well as the works~\cite{Yousaf2016b,Yousaf2017} for possible evolution of the stellar system. In the present work our plan is to study the entire system of gravastar in the background of modified teleparallel equivalent of general relativity (TEGR) which is well-known as $f(\mathbb{T})$ gravity and to observe different physical features of the gravitating system. Our previous successful projects on the study of compact stars~\cite{Das2015,Das2016} as well as gravastar~\cite{Das2017} and the work of Yousaf et al.~\cite{Yousaf2019b} motivate us to 
exploit the alternative formalism, i.e., the $f(\mathbb{T})$ gravity~\cite{Ferraro2007,Ferraro2008,Fiorini2009,Ferraro2009,Linder2010} to the case of the gravastar, a viable alternative to the ultimate stellar phase of a collapsing star, i.e., the black hole. Interestingly, in Ref.~\cite{Yousaf2019b} the authors have investigated the effects of electromagnetic field on the isotropic spherical gravastar model in $f(R,\mathcal{T})$ gravity. They found out singularity-free exact solutions for different regions of the gravastar. Several realistic characteristics of the model have been studied in the presence of the electromagnetic field through graphical representations and thus have arrived at the conclusion that the electric charge has a definite role in describing some features, viz., the proper length, energy contents, entropy and equation of state parameter of the stellar system. The authors have also explored the stable region of the charged gravastars.

Like all the other modified theories of gravity, instead of  changing the source side of the Einstein field equations the 
geometrical part has been changed by taking generalized functional form of the argument as the gravitational Lagrangian in the corresponding Einstein-Hilbert action. In the present case, it is taken as a function of torsion $\mathbb{T}$. Generally, the spin is held responsible to be the source of torsion, but there are other possibilities in which torsion emerges in different context. Most of the works under the background of $f(\mathbb{T})$ gravity have been done in cosmology as available in literature~\cite{Wu2010a,Tsyba2011,Dent2011,Chen2011,Bengochea2011,Wu2010b,Yang2011,Zhang2011,Li2011,Wu2011,Bamba2011}.

However, some astrophysical applications of $f(\mathbb{T})$ gravity can be observed in the following
works~\cite{Boehmer2011,Wang2011,Daouda2011,Nashed2013,Abbas2015a,Abbas2015b,Batti2017,Mai2017,Awad2017,Ganiou2017,Nashed2018,
saha2018,saha2019,sasa2018,Nashed2019,Awad2019,Capozziello2019,Bernard2020,Singh2019,chanda2019,debnath2019}. B{\"o}hmer et al.~\cite{Boehmer2011} examined the existence of relativistic stars in $f(\mathbb{T})$ gravity and explicitly constructed several classes of static perfect fluid solutions as well as the conservation equation for two different choices of tetrad. In the work~\cite{Wang2011} the author discussed spherically symmetric static solutions in $f(\mathbb{T})$ gravity theory including Maxwell term and found out explicitly the Reissner-Nordstr{\"o}m-(anti-)de Sitter solution as well as the Schwarzschild-(anti-)de Sitter solution. Also, new spherically symmetric solutions of black holes and wormholes were obtained with a constant torsion along with different forms of radial pressure by Daouda et al.~\cite{Daouda2011}. The author obtained a special spherically symmetric solution in $f(\mathbb{T})$ gravity theory by applying a non-diagonal spherically symmetric tetrad field in the work~\cite{Nashed2013}. Abbas et al.~\cite{Abbas2015a} investigated anisotropic compact stars specifically strange stars in $f(\mathbb{T})$ gravity under the Krori-Barua~\cite{KB1975} metric to a static spherically symmetric metric and found out that their strange star model is unstable whereas in another work~\cite{Abbas2015b} they studied the existence of charged strange stars in $f(\mathbb{T})$ gravity with MIT bag model. It is interesting to note that Batti et al.~\cite{Batti2017} have studied the dynamical instability of cylindrical compact objects in the framework of $f(\mathbb{T})$ gravity. A new class of the spherically-symmetric solutions have been obtained for the case of black holes and dark wormholes in $f(\mathbb{T})$ gravity by Mai et al.~\cite{Mai2017}. In work~\cite{Awad2017} authors provided the higher dimensional charged Anti-de-Sitter black hole solution considering a quadratic term in the functional form of $f(\mathbb{T})$. Ganiou et al.~\cite{Ganiou2017} studied the influence of a static very strong magnetic field on the neutron stars in $f(\mathbb{T})$ gravity and found out that the hadronic stars with very small hyperon contributions were compatible with the obtained results. Exact solutions for the charged anti-de Sitter BTZ black holes in Maxwell-$f(\mathbb{T})$ gravity were derived by Nashed and Capozziello~\cite{Nashed2018}. 

In their work~\cite{saha2018,saha2019} the authors studied anisotropic quintessence strange stars and  general compact stars in $f(\mathbb{T})$ gravity with modified Chaplygin gas respectively. In an interesting work~\cite{sasa2018} the authors investigated self-gravitating configurations of the polytropic fluid considering quadratic term in $f(\mathbb{T})$ gravity. On the other hand, several authors~\cite{Nashed2019,Awad2019,Capozziello2019,Bernard2020} have studied rotating AdS black holes in $f(\mathbb{T})$ gravity considering different physical situations. Singh et al.~\cite{Singh2019} examined a model of Einstein's cluster mimicking compact star under the background of $f(\mathbb{T})$ gravity whereas Chanda et al.~\cite{chanda2019} have studied anisotropic compact objects in $f(\mathbb{T})$ gravity in Finch-Skea geometry. In all these works available in literature one can notice that the works have been done in $f(\mathbb{T})$ gravity  either considering a model of compact stars or black holes or wormholes in different possible physical situations but not considering the model of gravastar except the work~\cite{debnath2019} where unlike our case the author have studied the charged gravastar in $f(\mathbb{T})$ gravity. 

Our work can be considered as a unique study where we have considered an isotropic model of uncharged gravastars which is more general and we have studied the effect of torsion in the interior as well as the exterior solution of the gravitating system. Gravastar can be thought of as an  alternative to the mysterious end state of gravitationally collapsing star, i.e., black hole which is believed to overcome any kind of repulsive nonthermal pressure of degenerate elementary particles. In our study we have obtained the interior solutions which are free from any central singularity unlike many of the other works in this field which consider a conformal symmetry in the geometry of the space-time. We have extensively studied the shell region which contains the stiff fluid and the total mass of the gravastar. We have explored different physical features, viz., the pressure-density, energy, entropy, proper length of the shell region analytically along with graphical representation unlike any other previous work in this field. From the graphical analysis it is clear that the pressure or density of the shell increases gradually with the thickness of the shell. This eventually suggests that the shell becomes more denser at the exterior than the interior boundary. We have found out the surface energy density and surface pressure at the boundary of the gravastar and arrived at the equation of state for boundary region. Interestingly, unlike the previous work in gravastar we have found out the exterior solution in $f(\mathbb{T})$ gravity which is turned out to be well-known Schwarzchild-de Sitter solution.

The outline of the present investigation is as follows : In Section~\ref{sec2} we provide the basic mathematical 
formalism of the $f(\mathbb{T})$ theory of gravity. Thereafter the field equations in $f(\mathbb{T})$ gravity
have been written assuming a specific form of gravitational Lagrangian, i.e., $f(\mathbb{T})$ in Section~\ref{sec3} whereas in Section~\ref{sec4}, the solutions of the field equations have been provided for different regions, i.e., interior, exterior and the shell of the gravastar. In Section~\ref{sec5} we discuss the junction conditions which are very important in connection to the three regions of the gravastar. Several physical features of the model, viz., the energy, entropy, proper length and equation of state have been discussed in Section~\ref{sec6}. Finally, in Sections~\ref{sec7} and \ref{sec8} we respectively discuss on the status of gravastar and pass some concluding remarks.

\section{Basic mathematical formalism of the $\bf{\textit{f}\left(\mathbb{T}\right)}$ theory}\label{sec2}
The action of $f(\mathbb{T})$ theory~\cite{Ferraro2009,Li2011,Boehmer2011,Daouda2011} is taken as (with geometrized units
$G=c=1$)
\begin{equation}
\mathbb{S}[e^i_{~\mu},~\phi_A ] = \int d^4x~e\left[ \frac{1}{16 \pi} f(\mathbb{T}) +
\mathcal{L}_{matter} (\phi_A)  \right]\label{eq1},
\end{equation}
where $\phi_A$ represents the matter fields and $f(\mathbb{T})$ is an arbitrary
analytic function of the torsion scalar $\mathbb{T}$. The torsion scalar is usually
constructed from the torsion and contorsion tensor as follows:
\begin{equation}
\mathbb{T}= S_\sigma^{\,~\mu\nu} \mathbb{T}^\sigma_{~\mu\nu},\label{eq2}
\end{equation}
where
\begin{equation}
\mathbb{T}^\sigma_{~\mu\nu} =  \tilde{\Gamma}^\sigma_{~\mu\nu}-\tilde{\Gamma}^\sigma_{~\nu\mu} 
= e_i^{~\sigma} \left( \partial_\mu e^i_{~\nu} - \partial_\nu
e^i_{~\mu} \right),\label{eq3}
\end{equation}

\begin{equation}
K^{\mu\nu}_{~~~\sigma} =  -\frac{1}{2} \left( \mathbb{T}^{\mu\nu}_{~~\sigma}-
\mathbb{T}^{\nu\mu}_{~~\sigma}-\mathbb{T}_{\sigma}^{~\mu\nu}\right),\label{eq4}
\end{equation}
are torsion and contorsion tensor respectively and new components of tensor
$S_\sigma^{~\mu\nu}$ can be written as 
\begin{equation}
S_\sigma^{~\mu\nu} =  \frac{1}{2} \left( K^{\mu\nu}_{~~~\sigma}+
\delta_\sigma^\mu \mathbb{T} ^{\beta\nu}_{~~\beta}-\delta_\sigma^\nu \mathbb{T}
^{\beta\mu }_{~~\beta} \right).\label{eq5}
\end{equation}

Here $e^i_{~\mu}$ are the tetrad fields by which we can define
any metric as $g_{\mu \nu} =  \eta_{ij} e^i_{~\mu} e^j_{~\nu}$ with 
$\eta_{ij} = diag(1,-1,-1,\\-1)$ and $e_i^{~\mu} e^i_{~\nu} =
\delta_\nu^\mu $,~$ e= \sqrt{-g} = det[e^i_{~\mu}]$.

Variation of the action (\ref{eq1}) with respect to the tetrad,
yield the field equations of $f(\mathbb{T})$ gravity~\cite{Ferraro2009,Li2011,Boehmer2011,Daouda2011} as 
\begin{eqnarray}
S_i^{~\mu \nu}f_{\mathbb{T}\mathbb{T}} \partial_\mu \mathbb{T} + e^{-1} \partial_\mu ( e
S_i^{~\mu \nu}) f_\mathbb{T} -\mathbb{T}^\sigma_{~\mu i}S_\sigma^{\,~\nu \mu} f_\mathbb{T}
+\frac{1}{4} e_i^{~\nu} f = 4 \pi T_i^\nu,\label{eq6}
\end{eqnarray}
where
\[ S_i^{~\mu \nu} =e_i^{~\sigma} S_\sigma^{~\mu \nu}, ~~f_\mathbb{T} =
\frac{\partial f}{\partial \mathbb{T}}~~~\&  ~~~f_{\mathbb{T}\mathbb{T}} = \frac{\partial^2
f}{\partial \mathbb{T}^2}.~~\]

In the above Eq. (\ref{eq6}) the symbol $T_i^\nu$ denotes the energy stress tensor of the perfect fluid which is given as
\begin{equation}
T_{\mu\nu}=(\rho+p)u_\mu u_\nu- p g_{\mu\nu},\label{eq7}
\end{equation}
with $ u^{\mu}u_{\mu} = 1,$ where $u_\mu$, $\rho$ and $p$ are respectively the four-velocity vectors, matter-energy density and fluid pressure of the system.

\section{The field equations in $\bf{\textit{f}\left(\mathbb{T}\right)}$ gravity}\label{sec3}
Let us assume that the space-time is spherically symmetric and static. This actually ensure that there is no translational or rotational movement taking place in the system under consideration. Hence, the line element can be taken as
\begin{equation}
ds^2 = e^{\nu}dt^2 - e^{\lambda}dr^2 - r^2 (d\theta^2+\sin^2\theta d\phi^2),\label{eq8}
\end{equation}
where the metric potentials ${\nu}$ and ${\lambda}$ are functions of radial coordinate $r$ only.

For the invariance of the line element under the Lorentz transformation the tetrad matrix $[e^i_\mu]$ can be defined as
\begin{equation}
[e^i_{~\mu}]=diag[e^{\nu(r)/2},~e^{\lambda(r)/2},~r,~r \sin\theta].\label{eq9}
\end{equation}

Hence, one can obtain $$e=det[e^i_{~\mu}]=e^{\frac{\nu +\lambda}{2}}r^2 \sin\theta. $$

Now, we can write the torsion scalar (\ref{eq2}) as
\begin{equation}
\mathbb{T}(r) = \frac{2 e^{-\lambda}}{r}\left(\nu'+\frac{1}{r}\right).\label{eq10}
\end{equation}

In the above mathematical expression the prime $(')$ denotes derivative with respect to the radial coordinate $r$ and this notation will be followed afterwards also.

Inserting the components of $S_i^{~\mu \nu}$ and $\mathbb{T}^i_{~\mu\nu}$ in Eq. (\ref{eq6})
one can obtain the field equations as 
\begin{eqnarray}
4 \pi \rho = -\frac{e^{-\lambda}}{r} \mathbb{T}'
f_{\mathbb{T}\mathbb{T}}-\left[\frac{\mathbb{T}}{2}-\frac{1}{2
r^2}-\frac{e^{-\lambda}}{2r}\left(\nu'+\lambda'\right)\right]f_\mathbb{T}+\frac{f}{4} , \label{eq11}
\end{eqnarray}

\begin{equation}
4 \pi p =
\left[\frac{\mathbb{T}}{2}-\frac{1}{2r^2}\right]f_\mathbb{T}-\frac{f}{4},\label{eq12}
\end{equation}

\begin{eqnarray}
4 \pi p = \frac{e^{-\lambda}}{2}
\left(\frac{\nu'}{2}+\frac{1}{r}\right)\mathbb{T}' f_{\mathbb{T}\mathbb{T}}+\frac{\mathbb{T}}{4}f_\mathbb{T}
\nonumber\\+\frac{e^{-\lambda}}{2}\left[\frac{\nu''}{2}+\left(\frac{\nu'}{4}+\frac{1}{2r}\right)(\nu'-\lambda')\right]f_\mathbb{T}-\frac{f}{4},\label{eq13}
\end{eqnarray}

\begin{equation}
\frac{e^{-\lambda/2}cot{\theta}}{2r^2}\mathbb{T}'f_{\mathbb{T}\mathbb{T}}=0.\label{eq14}
\end{equation}

Also, the conservation equation in $f(\mathbb{T})$ theory~\cite{Boehmer2011} for spherically symmetric static spacetime reads as
\begin{equation}
4\pi p'+2\pi \nu'(\rho+p)=-\frac{\mathbb{T}'}{2r^2}f_{\mathbb{T}\mathbb{T}}.\label{eq15}
\end{equation}

Let us now consider the functional form of $f(\mathbb{T})$ as
\begin{equation}
f(\mathbb{T}) =a\mathbb{T}+b, \label{eq16}
\end{equation}
which is consistent with Eq. (\ref{eq14}). Here $a$ and $b$ are constants which consequently implies that $f_\mathbb{T}=a$ and $f_{\mathbb{T}\mathbb{T}}=0$.

For the above mentioned form of $f(\mathbb{T})$ along with Eq. (\ref{eq10}) the Eqs. (\ref{eq11})-(\ref{eq13}) can be rewritten as
\begin{equation}
4 \pi\rho = \frac{a}{2r}\left[\lambda'e^{-\lambda}-\frac{e^{-\lambda}}{r}+\frac{1}{r}\right]+\frac{b}{4} ,\label{eq17}
\end{equation}

\begin{equation}
4 \pi p = \frac{a}{2r}\left[\nu'e^{-\lambda}+\frac{e^{-\lambda}}{r}-\frac{1}{r}\right]-\frac{b}{4},\label{eq18}
\end{equation}

\begin{eqnarray}
4 \pi p =\frac{ae^{-\lambda}}{2}\left[\frac{\nu''}{2}+\left(\frac{\nu'}{4}+\frac{1}{2r}\right)\left(\nu'-\lambda'\right)\right]- \frac{b}{4}.\nonumber \\ \label{eq19}
\end{eqnarray}

Also, the conservation equation (\ref{eq15}) in $f(\mathbb{T})$ gravity takes the form as
\begin{equation}
\frac{dp}{dr}+\frac{\nu'}{2}(\rho+p)=0.\label{eq20}
\end{equation}

From the above equation it is evident that for the equilibrium condition of a 
gravitating system the pressure gradient which can be termed as the hydrostatic 
force must be counterbalanced by the gravitational force.

\section{The solutions of the field equations for different regions of gravastar}\label{sec4}

\subsection{Interior space-time}
As per the proposition of Mazur-Mottola~\cite{Mazur2001,Mazur2004}, we have assumed the
equation of state (EOS) for the interior region as
\begin{equation}
p=-\rho.\label{eq21}
\end{equation}

The above EOS is well-known as the dark energy EOS with the parametric value $-1$.

Again from the above EOS as well as Eq. (\ref{eq20}) one can obtain
\begin{equation}
\rho = \rho_0~(constant),\label{eq22}
\end{equation}
and the pressure can be found out to be
\begin{equation}
p=-\rho_0.\label{eq23}
\end{equation}

Now, from Eqs. (\ref{eq17}) and (\ref{eq22}) one can find the metric potential $\lambda$ as
\begin{equation}
e^{-\lambda}= 1-\left(\frac{8\pi\rho_0}{3a}-\frac{b}{6a}\right)r^2+\frac{k_1}{r},\label{eq24}
\end{equation}
where $k_1$ is an integration constant which is set to zero due to regularity of the solution at the
center ($r=0$). Hence one can write
\begin{equation}
e^{-\lambda}= 1-\left(\frac{8\pi\rho_0}{3a}-\frac{b}{6a}\right)r^2.\label{eq25}
\end{equation}

Again from Eqs. (\ref{eq17}), (\ref{eq18}), (\ref{eq22}) and (\ref{eq23}) we get the
following relation between the metric potentials $\nu$ and $\lambda$ as
\begin{equation}
e^{\nu}=k_2e^{-\lambda},\label{eq26}
\end{equation}
where $k_2$ is an integration constant. Here, the space-time is found to be centrally non-singular.

Also the gravitational mass $M(D)$ can be written as
\begin{equation}
M(D)= \int_0^{r_1=D} 4\pi r^2{\rho_0} dr=\frac{4}{3}\pi
D^3\rho_0.\label{eq27}
\end{equation}

\subsection{Shell}
The shell consists of ultrarelativistic fluid and it obeys the EOS $p=\rho$.
In connection to cold baryonic universe Zel'dovich~\cite{zeldovich1972} first 
conceived the idea of this kind of fluid and it is also known as the stiff fluid.
In the present case we can argue that this may arise from the thermal
excitations with negligible chemical potential or from conserved number density of the
gravitational quanta at the zero temperature~\cite{Mazur2001,Mazur2004}. Several authors have
extensively exploited this type of fluid to explore various cosmological~\cite{Madsen1992}
as well as astrophysical~\cite{carr1975,wesson1978,braje2002,linares2004} phenomena.

Within the non-vacuum region, i.e., the shell one can observe that it
is very difficult to find solution of the field equations. However, within 
the framework of the thin shell limit, i.e., $0< e^{-\lambda}\ll1$ it is possible 
to find an analytical solution. As prescribed by Israel~\cite{Israel1966} 
we can possibly argue that the intermediate region between the two space-times 
(in this case the vacuum interior and the Schwarzschild exterior) must be a thin shell. 
Also within the thin shell region any parameter which is a function of $r$ is, in general, $\ll1$ as $r\rightarrow 0$. 
Due to this kind of approximation along with the above EOS as well as Eqs. (\ref{eq17}), (\ref{eq18}) and (\ref{eq19}), 
one can obtain the following equations
\begin{equation}\label{eq28}
\frac{de^{-\lambda}}{dr}=\frac{2}{r}+\frac{br}{a},
\end{equation}

\begin{equation}\label{eq29}
\left(\frac{a\nu'}{8}+\frac{3a}{4r}\right)\frac{de^{-\lambda}}{dr}=\frac{a}{2r^2}+\frac{b}{2}.
\end{equation}

Integrating Eq. (\ref{eq28}) we get
\begin{equation}\label{eq30}
e^{-\lambda}= 2\ln r + \frac{br^2}{2a}+k_3,
\end{equation}
where $k_3$ is an integration constant and $r$ lies within the range
$D\leq\,r\leq{D+\epsilon}$. Also we get $k_3\ll1$ under the condition $\epsilon\ll1$ as well as $e^{-\lambda}\ll1$.

Again from Eqs. (\ref{eq28}) and (\ref{eq29}) one can obtain
\begin{equation}
e^{\nu}=k_4\left(\frac{br^2+2a}{r^4}\right),\label{eq31}
\end{equation}
where $k_4$ is an integration constant.

Also from Eqs. (\ref{eq20}) and (\ref{eq31}) along with the EOS $p=\rho$, one gets 
\begin{equation}\label{32}
p=\rho=k_5\left(\frac{r^4}{br^2+2a}\right),
\end{equation}
where $k_5$ is a constant.

\begin{figure}[h]
\centering
\includegraphics [width=0.5\textwidth]{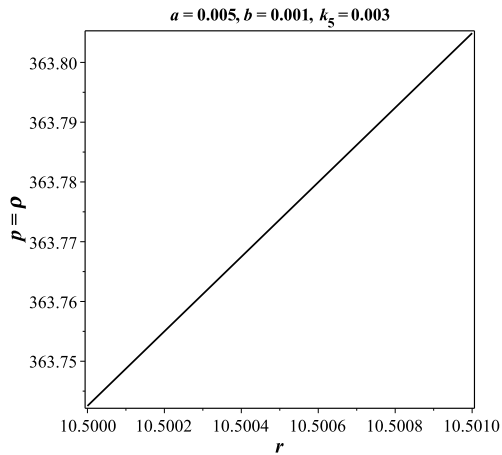}
\caption{Pressure $p=\rho$ (km$^{-2}$) of the ultrarelativistic
fluid in the shell is plotted with respect to the radial
coordinate $r$ (km)}
\end{figure}

\subsection{Exterior space-time}
The exterior region obeys the EOS $p=\rho=0$. Therefore, for the above EOS from Eq. (\ref{eq17}) we can find 
\begin{equation}
e^{-\lambda}=\left(1+\frac{c'}{r}+\frac{br^2}{6a}\right),\label{eq33}
\end{equation}
where $c'$ is an integration constant.

Again, using Eqs. (\ref{eq17}) and (\ref{eq18}) we get,
\begin{equation}
\nu'+\lambda'=constant.\label{eq34}
\end{equation}

By taking a suitable transformation in the time co-ordinate we can set the above equation as
\begin{equation}
\nu'+\lambda'=0\label{eq35}
\end{equation}

and we get
\begin{eqnarray}
e^\nu=e^{-\lambda}=\left(1+\frac{c'}{r}+\frac{br^2}{6a}\right).\label{eq36}
\end{eqnarray}

Comparing the solution as given by Eq. (36) with the well-known Schwarzschild-de Sitter metric, we can set 
$c'=-2M$ and $\frac{b}{6a}=-\frac{\Lambda}{3}$ so that the line element for the exterior region
can be written as
\begin{eqnarray}
&\qquad \hspace{-0.5cm}ds^2=\left(1-\frac{2M}{r}-\frac{\Lambda}{3}r^2\right)dt^2-\left(1-\frac{2M}{r}-\frac{\Lambda}{3}r^2\right)^{-1}dr^2\nonumber\\
&\qquad -r^2\left(d\theta^2+\sin^2\theta d\phi^2\right),\label{eq37}
\end{eqnarray}
where $M$ is the total mass and $\Lambda$ is the cosmological constant.

\section{Junction condition}\label{sec5}
We have already mentioned that the gravastar has three regions, i.e., the interior region (A), shell (B), and exterior region (C). The interior region (A) is connected with the exterior region (C) through the shell. There should be a smooth matching at the junction interface between the regions (A) and (C) of the gravastar if one consider the Darmois-Israel formalism~\cite{Israel1966,Darmois1927}. Hence, the metric coefficients are continuous at the junction surface ($\Sigma$), i.e. at $r=D$, though their derivatives may be discontinuous. However, exploiting the above mentioned formalism one can easily determine the surface stress-energy $\mathscr{S}_{ij}$.

The intrinsic surface stress-energy tensor $\mathscr{S}_{ij}$ is written by the Lanczos equation~\cite{lanczos1924,sen1924,Darmois1927,Israel1966,perry1992,lake1996} as
\begin{equation}
\mathscr{S}^i_j=-\frac{1}{8\pi}(\kappa^i_j-\delta^i_j\kappa^k_k),\label{eq38}
\end{equation}
where $\kappa_{ij}=K^+_{ij}-K^-_{ij}$ present the discontinuity in
the second fundamental forms or extrinsic curvatures. Here the
signs ``$+$'' and ``$-$'' correspond to the interior and the
exterior regions respectively. Moreover, the second fundamental
forms~\cite{rahaman2006,rahaman2009,usmani2010,rahaman2010,dias2010,rahaman2011}
associated with the two sides of the shell can be written as 
\begin{equation}
K_{ij}^{\pm}=-n_{\nu}^{\pm}\left[\frac{\partial^{2}x_{\nu}}{\partial
\zeta^{i}\partial\zeta^{j}}+\Gamma_{\alpha\beta}^{\nu}\frac{\partial
x^{\alpha}}{\partial \zeta^{i}}\frac{\partial x^{\beta}}{\partial
\zeta^{j}} \right]|_\Sigma, \label{eq39}
\end{equation}
where $\zeta^{i}$ denote the intrinsic coordinates on the shell, $n_{\nu}^{\pm}$ being the unit normals to the
surface $\Sigma$ and for the spherically symmetric static metric
\begin{equation}
ds^2=f(r)dt^2-\frac{dr^2}{f(r)}-r^2(d\theta^2+\sin^2\theta\,d\phi^2),\label{eq40}
\end{equation}
one can write $n_{\nu}^{\pm}$  as
\begin{equation}
n_{\nu}^{\pm}=\pm\left|g^{\alpha\beta}\frac{\partial f}{\partial
x^{\alpha}}\frac{\partial f}{\partial x^{\beta}}
\right|^{-\frac{1}{2}}\frac{\partial f}{\partial x^{\nu}},\label{eq41}
\end{equation}
with $n^{\mu}n_{\mu}=1$.

The surface stress-energy tensor can be defined as {\bf $\mathscr{S}_{ij}=diag  [{\sigma,-\upsilon,-\upsilon,-\upsilon }$]} 
using the Lanczos equation, where $\sigma$ is the surface energy density and $\upsilon$ is the surface pressure. The surface energy density
($\sigma$) and the surface pressure ($\upsilon$) can be respectively given by
\begin{equation}
\sigma=-\frac{1}{4\pi D}\left[\sqrt{f}\right]^+_-, \label{eq42}
\end{equation}

\begin{equation}
\upsilon=-\frac{\sigma}{2}+\frac{1}{16\pi}\left[\frac{f^{'}}{\sqrt{f}}\right]^+_-.\label{eq43}
\end{equation}

Using the above two equations and putting $\Lambda=-\frac{b}{2a}$ we therefore obtain
\begin{equation}
\sigma=-\frac{1}{4\pi D}\left[\sqrt{1-\frac{2M}{D}+\frac{bD^2}{6a}}-\sqrt{1-\left(\frac{8\pi\rho_0}{3a}-\frac{b}{6a}\right)D^2}\right],\label{eq44}
\end{equation}

\begin{equation}
\upsilon=\frac{1}{8\pi D}\left[\frac{(1-\frac{M}{D}+\frac{bD^2}{3a})}{\sqrt{1-\frac{2M}{D}+\frac{bD^2}{6a}}}-\frac{1-\frac{16\pi\rho_0D^2}{3a}+\frac{bD^2}{3a}}{\sqrt{1-\left(\frac{8\pi\rho_0}{3a}-\frac{b}{6a}\right)D^2}}\right]. \label{eq45}
\end{equation}

Moreover, one can write the mass of the thin shell as
{\small{
\begin{eqnarray}
& \qquad\hspace{-13cm}m_s=4\pi D^2\sigma \nonumber \\
\hspace{0.5cm}=D\left[\sqrt{1-\left(\frac{8\pi\rho_0}{3a}-\frac{b}{6a}\right)D^2}-\sqrt{1-\frac{2M}{D}+\frac{bD^2}{6a}}\right],\label{eq46}
\end{eqnarray}}}
where $M$ is the total mass of the gravastar and is given by
{\small{
\begin{equation}\label{eq47}
M=\frac{4\pi\rho_0D^3}{3a} 
+ m_s \sqrt{1-\left(\frac{8\pi\rho_0}{3a}-\frac{b}{6a}\right)D^2}-\frac{m_s^2}{2D}.
\end{equation}}}

\section{Physical features of gravastar model}\label{sec6}

\subsection{Energy content}
In the interior region as we consider the EOS, $p=-\rho$, one can argue that this indicates the negative energy
region confirming the repulsive nature of the core. However, the energy within the shell can be found out to be
\begin{eqnarray}\label{eq48}
&\qquad\hspace{-1cm}\mathcal{E}=\int_{D}^{D+\epsilon}4\pi\rho\,r^{2}dr=4\pi k_5\int_{D}^{D+\epsilon}\frac{r^6}{\left(br^2+2a\right)}dr\nonumber \\
&\qquad\hspace{-0.5cm}=4\pi k_5\left[\frac{r^5}{5b}-\frac{2ar^3}{3b^2}+\frac{4a^2r}{b^3}-\frac{4\sqrt{2}a^{5/2}{\arctan}\left(r\sqrt{\frac{b}{2a}}\right)}{b^{7/2}}\right]_D^{D+\epsilon}.\nonumber \\  
\end{eqnarray}

\begin{figure}[h]
\centering
\includegraphics [width=0.49\textwidth]{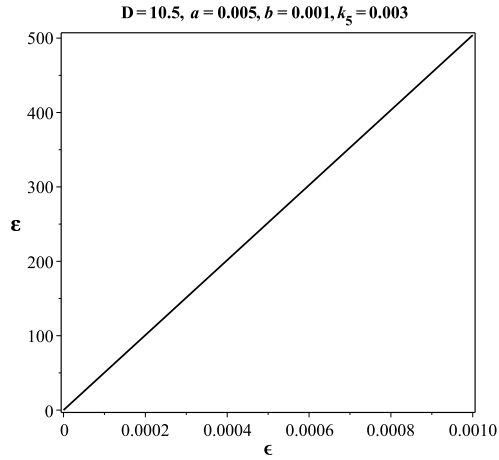}
\caption{Energy $\varepsilon$ (km) within the shell is plotted
with respect to the thickness of the shell $\epsilon$ (km)}
\end{figure}

\subsection{Entropy}
Following the prescription of Mazur and Mottola~\cite{Mazur2001,Mazur2004} one can argue that in the interior region (A)
the entropy density is zero which is consistent with a single condensate state. Though the entropy within the shell can be written as
\begin{equation}
\mathcal{S}=\int_{D}^{D+\epsilon}4\pi r^{2}s(r)\sqrt{e^{\lambda}}dr,\label{eq49}
\end{equation}
where $s(r)$ is the entropy density for local temperature $T(r)$ and can be written as~\cite{Mazur2001,Mazur2004}
\begin{equation}
s(r)=\frac{\alpha^2k_B^2T(r)}{4\pi\hbar^2 } =
\alpha\left(\frac{k_B}{\hbar}\right)\sqrt{\frac{p}{2 \pi
}},\label{eq50}
\end{equation}  
where $\alpha$ is a dimensionless constant. Here we have assumed the Planckian units, i.e., $k_B=\hbar=1$ along with the geometrized units, i.e., $G=c=1$ as mentioned earlier. So, the entropy density within the shell is given by
\begin{equation}
s(r)=\alpha\sqrt{\frac{p}{2\pi}}.\label{eq51}
\end{equation}

Therefore, Eq. (\ref{eq45}) can be written as
\begin{equation}\label{eq52}
\mathcal{S}=\sqrt{8\pi\,k_5}~\alpha\int_{D}^{D+\epsilon}\frac{r^4}{\sqrt{\left(br^2+2a \right)\left( 2\ln r+\frac{br^2}{2a}+k_3\right)}}dr.
\end{equation}

Integrating the above equation one can obtain
\begin{eqnarray}\label{eq53}
&\qquad\hspace{-1cm}\mathcal{S}= \left( {\frac {2\alpha\,{r}^{5}\sqrt {\pi k_5 }}{5{a}^{5/2}{k_{
{3}}}^{3/2}}} \right) \Big[  \left( {\frac {5
\,ab{r}^{2}}{28}}-{a}^{2} \right) \ln r  + \left( k_{{3
}}+\frac{1}{5}\right) {a}^{2}-{\frac {5\,ab{r}^{2}}{28} \left( k_{{3}}+{
\frac {8}{7}} \right) }+{\frac {5\,{b}^{2}{r}^{4}}{144}} \Big]^{D+\epsilon}_D.
\end{eqnarray}

\begin{figure}[h]
\centering
\includegraphics [width=0.48\textwidth]{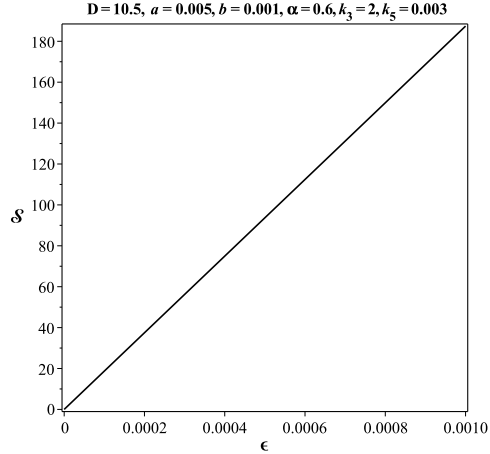}
\caption{Entropy $\mathcal{S}$ within the shell is plotted with respect to
the thickness of the shell $\epsilon$ (km)}
\end{figure}

\subsection{Proper length of the shell}
As per the conjecture of Mazur and Mottola~\cite{Mazur2001,Mazur2004} the stiff fluid 
shell which is situated at the surface $r=D$ defines
the phase boundary of region (A) and the proper thickness of the shell can be assumed to be
very small, i.e., $\epsilon\ll1$. Hence, the lower bound of region (C) is 
$r={D+\epsilon}$. Therefore, one can write the proper thickness between two interfaces, i.e., of the shell as
\begin{equation}
\ell= \int_D^{D+\epsilon}
\sqrt{e^{\lambda}}dr=\int_D^{D+\epsilon}\frac{dr}{\sqrt{2\ln r +\frac{br^2}{2a}+k_3}}.\label{eq54}
\end{equation}

Integrating the above equation one can obtain
{\small{
\begin{eqnarray}\label{eq55}
\ell=\frac{1}{k_3^{3/2}}\Big[ (D+\epsilon)\ln \left( \frac{1}{D+\epsilon}\right)+D\ln D-\frac{Db\epsilon}{4a}(D+\epsilon)
-\frac{b\epsilon^3}{12a}+\epsilon(k_3+1)\Big].  
\end{eqnarray}}}

\begin{figure}[h]
\centering
\includegraphics [width=0.5\textwidth]{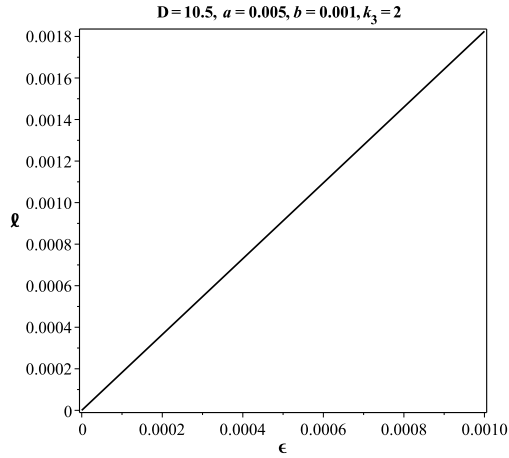}
\caption{Proper length $\ell$ (km) of the shell is plotted with
respect to the thickness of the shell $\epsilon$ (km).}
\end{figure}

\subsection{Equation of state}
We can express the EOS at $r=D$ as 
\begin{equation}
\upsilon=\omega(D)\sigma. \label{eq56}
\end{equation}

Hence, from Eqs. (\ref{eq44}) and (\ref{eq45}) one can explicitly write the
equation of state parameter as
\begin{equation}
\omega(D)=\frac{\left[\frac{(1-\frac{M}{D}+\frac{bD^2}{3a})}{\sqrt{1-\frac{2M}{D}+\frac{bD^2}{6a}}}-\frac{1-\frac{16\pi\rho_0D^2}{3a}+\frac{bD^2}{3a}}{\sqrt{1-\left(\frac{8\pi\rho_0}{3a}-\frac{b}{6a}\right)D^2}}\right]}{2\left[\sqrt{1-\left(\frac{8\pi\rho_0}{3a}-\frac{b}{6a}\right)D^2}-\sqrt{1-\frac{2M}{D}+\frac{bD^2}{6a}}\right]}.\label{eq57}
\end{equation}

\section{Discussion on the status of the gravastar model}\label{sec7}
For finding out some possibility to verify our results of the model regarding status 
as well as detection of the gravastar under $f(\mathbb{T})$ gravity one may either 
consider the gravitational wave signatures~\cite{Pani2009,Pani2010a,Pani2010b,Abbott2016,Cardoso2016,Chirenti2016,Bamba2013} 
of gravastars or study their lensing effects as suggested by many authors, solely for gravastars~\cite{Kubo2016}
and for $f(\mathbb{T})$ gravity~\cite{Ruggiero2016}. Pani et al.~\cite{Pani2009,Pani2010a,Pani2010b} 
illustrated how the presence or absence of an event horizon could produce qualitative differences in the gravitational waves emitted by ultracompact objects under different physical aspects and for ultracompact object with no event horizon, they considered
a non-rotating thin shell gravastar.

Recently Cardoso et al.~\cite{Cardoso2016} have raised a possibility that the measured ringdown signal of $GW150914$~\cite{Abbott2016} which has been recently detected by the interferometric LIGO detectors may be due to horizonless objects such as gravastars etc. among many others. In another work Chirenti and Rezzolla~\cite{Chirenti2016} have concluded that it is not possible to model the measured ringdown signal of $GW150914$ as due to gravastar in spite of having very limited knowledge of the perturbative response of rotating gravastar. 

Furthermore, in a work Bamba et al.~\cite{Bamba2013} explored the possibility 
of further gravitational wave modes in $f(\mathbb{T})$ gravity 
and explicitly demonstrated that gravitational wave modes are equivalent to those 
in general relativity, though they achieved their result  calculating the Minkowskian limit 
for a specific class of analytic function of  $f(\mathbb{T})$. Interestingly, ``Was 
gravitational wave signal from a gravastar, not black holes?'' is now a valid question
in the arena of scientific community~\cite{Aron2016}.

However, for the possible detection of gravastar due to their lensing effect one may adopt 
the methodology suggested by Kubo and Sakai~\cite{Kubo2016} based on the gravastar model 
developed by Visser and Wiltshire~\cite{Visser2004}. 
Now, assuming its surface to be optically transparent they have calculated the image 
of a companion which has been supposed to rotate around the gravastar and have 
found that some characteristic images have appeared, depending on whether the 
gravastar possess unstable circular orbits of photons (Model 1) or not (Model 2).
For Model 2, Kubo and Sakai have calculated the total luminosity change, which 
has been called microlensing effects. Here the maximal luminosity 
could have been considerably larger than the black hole with the same mass.
One can study the similar effects under $f(\mathbb{T})$ gravity to 
compare the effects of modified gravity on the above mentioned
tests with that of the results based on general theory of relativity.

\section{Conclusion}\label{sec8}
In this paper we have presented a unique stellar model of gravastar under
the $f(\mathbb{T})$ gravity following the model proposed by
Mazur-Mottola~\cite{Mazur2001,Mazur2004} in the framework of
general relativity. Mazur-Mottola~\cite{Mazur2001,Mazur2004} 
described the spherically symmetric stellar structure of gravastar 
by the three different regions: interior core region, intermediate 
thin shell region, and exterior spherical
region with specific EOS for each of the region. With this type
of specifications for the stellar system we have found out a set of exact and
singularity-free solutions of gravastar which represents 
several properties which are interesting as well as physically
viable within the $f(\mathbb{T})$ theory of alternative gravity.

We have examined and explained several salient aspects of 
the solution set based upon the above mentioned structural
form of a gravastar and those can be summarized below:\\

\indent (1) {\it Pressure-density profile}:\\ 
\indent (i) In the core region, the density turns out 
to be constant and hence the pressure is eventually
negative in nature maintaining a constant value.\\
\indent (ii) The variation of pressure or density (as $p=\rho$) 
of the ultrarelativistic fluid in the shell with respect 
to the radial coordinate $r$ is shown in Fig. 1 which 
reveals that it increases gradually with the thickness of the shell. 
This eventually suggests that the shell becomes more denser at the
exterior boundary than the interior boundary.

(2) {\it Energy content}:\\ 
\indent (i) The interior region is governed by the EOS, $p=-\rho$, which 
indicates the negative energy region confirming the repulsive nature of the core.\\
\indent (ii) The energy within the shell when plotted with respect to the thickness of 
the shell $\epsilon$ (in Fig. 2) shows an increasing profile. 

(3) {\it Entropy}:\\
\indent (i) In the interior region (A) the entropy density is zero which is consistent
with a single condensate state of the core.\\ 
\indent (ii) Within the shell the entropy $\mathcal{S}$ has been
plotted with respect to the thickness of the shell $\epsilon$ (in Fig. 3). 
From the plot it has been revealed that the entropy 
is gradually increasing with respect to the thickness of
the shell $\epsilon$ which is a physically valid feature as well as 
suggesting a maximum value on the surface of the gravastar.

(4) {\it Proper length}: The proper length $\ell$ of the shell 
also shows a gradual increasing profile (in Fig. 4) which is plotted 
with respect to the thickness of the shell $\epsilon$.

(5) {\it Equation of state}: We find out the equation of state parameter $\omega(D)$
at the junction interface of the shell and the vacuum exterior and this parameter has 
an important physical significance.

Unlike Einstein's general relativity there are different terms involving the constant $a$ and $b$ 
in the expressions of different physical parameters of the model due to the inclusion of function of
torsion scalar $\mathbb{T}$ which is assumed to have the form $f(\mathbb{T})=a\mathbb{T}+b$
as gravitational Lagrangian in the corresponding action. This modification has a definite role 
and makes the differences between the expressions in both the theories and can be tested by 
executing a comparative study between the present work and that of Rahaman et al.~\cite{Rahaman2015} 
and Ghosh et al.~\cite{ghosh2017} under 4-dimensional background.

In one of our earlier works~\cite{Das2015} we made a comment on an interesting work 
on the $f(\mathbb{T})$ gravity in connection to Krori and Barua metric~\cite{KB1975} as done
by Abbas et al.~\cite{Abbas2015a}. As a forthcoming work, we can therefore find out 
the possibility to study gravastar within the present formalism using Krori and Barua metric.

\section*{Acknowledgments} FR and SR are thankful for the support from the Inter-University Centre for Astronomy and Astrophysics (IUCAA), Pune, India 
for providing the Visiting Associateship under which a part of this work was carried out. FR is thankful to DST-SERB (EMR/2016/000193), Govt. of India for providing financial support.


\begin{thebibliography}{99}

\bibitem{Mazur2001} P. Mazur, E. Mottola, Report number: LA-UR-01-5067 (2001), arXiv:gr-qc/0109035.

\bibitem{Mazur2004} P. Mazur, E. Mottola, Proc. Natl. Acad. Sci. USA 101 (2004) 9545, arXiv:gr-qc/0407075. 

\bibitem{zeldovich1972} Y.B. Zel'dovich, Mon. Not. R. Astron. Soc. 160 (1972) 1P. 

\bibitem{Visser2004}  M. Visser, D.L. Wiltshire, Class. Quantum Grav. 21 (2004) 1135, arXiv:gr-qc/0310107. 

\bibitem{Cattoen2005} C. Cattoen, T. Faber, M. Visser, Class. Quantum Grav. 22 (2005) 4189, arXiv:gr-qc/0505137. 

\bibitem{Carter2005} B.M.N. Carter, Class. Quantum Grav. 22 (2005) 4551, arXiv:gr-qc/0509087.  

\bibitem{Bilic2006}  N. Bili\'{c}, G.B. Tupper, R.D. Viollier, J. Cosmol. Astropart. Phys. 02 (2006) 013, arXiv:astro-ph/0503427.  

\bibitem{Lobo2006} F.S.N. Lobo, Class. Quantum Grav. 23 (2006) 1525, arXiv:gr-qc/0508115.  

\bibitem{DeBenedictis2006} A. DeBenedictis, D. Horvat, S. Iliji\'{c}, S. Kloster, K.S. Viswanathan, Class. Quantum Grav. 23 (2006) 2303, arXiv:gr-qc/0511097. 

\bibitem{Lobo2007}  F.S.N. Lobo, A.V.B. Arellano, Class. Quantum Grav. 24 (2007) 1069, arXiv:gr-qc/0611083. 

\bibitem{Horvat2007}  D. Horvat, S. Iliji\'{c}, Class. Quantum Grav. 24 (2007) 5637, arXiv:0707.1636 [gr-qc]. 

\bibitem{Cecilia2007} C.B.M.H. Chirenti, L. Rezzolla, Class. Quantum Grav. 24 (2007) 4191, arXiv:0706.1513 [gr-qc]. 

\bibitem{Rocha2008} P. Rocha, R. Chan, M.F.A. da Silva, A. Wang, J. Cosmol. Astropart. Phys. 11 (2008) 010, arXiv:0809.4879 [gr-qc]. 

\bibitem{Horvat2008} D. Horvat, S. Iliji\'{c}, A. Marunovic, Class. Quantum Grav. 26 (2009) 025003, arXiv:0807.2051 [gr-qc]. 

\bibitem{Nandi2009} K.K. Nandi, Y.Z. Zhang, R.G. Cai, A. Panchenko, Phys. Rev. D 79 (2009) 024011, arXiv:0809.4143 [gr-qc]. 

\bibitem{Turimov2009} B.V. Turimov, B.J. Ahmedov, A.A. Abdujabbarov, Mod. Phys. Lett. A 24 (2009) 733, arXiv:0902.0217 [gr-qc]. 

\bibitem{Usmani2011} A.A. Usmani, F. Rahaman, S. Ray, K.K. Nandi, P.K.F. Kuhfittig, Sk.A. Rakib, Z. Hasan, Phys. Lett. B  701 (2011) 388, arXiv:1012.5605 [gr-qc]. 

\bibitem{Lobo2013}  F.S.N. Lobo, R. Garattini, J. High Energy Phys. 12 (2013) 065, arXiv:1004.2520 [gr-qc]. 

\bibitem{Bhar2014} P. Bhar, Astrophys. Space Sci. 354 (2014) 2109. 

\bibitem{Rahaman2015} F. Rahaman, S. Chakraborty, S. Ray, A.A. Usmani, S. Islam, Int. J. Theor. Phys. 54 (2015) 50, arXiv:1209.6291 [physics.gen-ph]. 

\bibitem{Ri1998} A.G. Riess et al., Supernova Search Team collaboration, Astron. J. 116 (1998) 1009, arXiv:astro-ph/9805201. 

\bibitem{Perl1999} S. Perlmutter et al., Supernova Cosmology Project collaboration, Astrophys. J. 517 (1999) 565, arXiv:astro-ph/9812133. 

\bibitem{Bern2000} P. de Bernardis et al., Nature 404 (2000) 955, arXiv:astro-ph/0004404. 

\bibitem{Hanany2000} S. Hanany et al., Astrophys. J. 545 (2000) L5, arXiv:astro-ph/0005123. 

\bibitem{Peebles2003} P.J.E. Peebles, B. Ratra, Rev. Mod. Phys. 75 (2003) 559, arXiv:astro-ph/0207347. 
 
\bibitem{Paddy2003} T. Padmanabhan, Phys. Rep. 380 (2003) 235, arXiv:hep-th/0212290. 

\bibitem{Clifton2012} T. Clifton, P.G. Ferreira, A. Padilla, C. Skordis, 513 (2012) 1, arXiv:1106.2476 [astro-ph.CO].

\bibitem{Yousaf2016a} Z. Yousaf, K. Bamba, M.Z.H. Bhatti, Phys. Rev. D 93 (2016) 124048.

\bibitem{Yousaf2018} Z. Yousaf, Astrophys. Space Sci. 363 (2018) 226.

\bibitem{Yousaf2019a} Z. Yousaf, Eur. Phys. J. Plus 134 (2019) 245, arXiv:2002.01316 [gr-qc].

\bibitem{Yousaf2016b} Z. Yousaf, K. Bamba, M.Z.H. Bhatti, Phys. Rev. D  93 (2016) 064059, arXiv:1603.02175 [gr-qc].

\bibitem{Yousaf2017} Z. Yousaf, Eur. Phys. J. Plus 132 (2017) 71. 

\bibitem{Das2015} A. Das, F. Rahaman, B.K. Guha, S. Ray, Astrophys. Space Sci. 358 (2015) 36, arXiv:1507.04959 [gr-qc]. 

\bibitem{Das2016} A. Das, F. Rahaman, B.K. Guha, S. Ray, Eur. Phys. J. C 76 (2016) 654, arXiv:1608.00566 [gr-qc]. 

\bibitem{Das2017} A. Das, S. Ghosh, B.K. Guha, S. Das, F. Rahaman, S. Ray, Phys. Rev. D 95 (2017) 124011, arXiv:1702.08873 [gr-qc]. 

\bibitem{Yousaf2019b} Z. Yousaf, K. Bamba, M. Z. Bhatti, U. Ghafoor, Phys. Rev. D 100 (2019) 024062, arXiv:1907.05233.

\bibitem{Ferraro2007} R. Ferraro, F. Fiorini, Phys. Rev. D 75 (2007) 084031, arXiv:gr-qc/0610067. 

\bibitem{Ferraro2008} R. Ferraro, F. Fiorini, Phys. Rev. D 78 (2008) 124019, arXiv:0812.1981 [gr-qc]. 

\bibitem{Fiorini2009} F. Fiorini, R. Ferraro, Int. J. Mod. Phys. A 24 (2009) 1686, arXiv:0904.1767 [gr-qc]. 

\bibitem{Ferraro2009} G.R. Bengochea, R. Ferraro, Phys. Rev. D 79 (2009) 124019, arXiv:0812.1205 [astro-ph]. 

\bibitem{Linder2010} E.V. Linder, Phys. Rev. D 81 (2010) 127301, Erratum: 82 (2010) 109902, arXiv :1005.3039 [astro-ph.CO].

\bibitem{Wu2010a} P. Wu, H. Yu, Phys. Lett. B 692 (2010) 176, arXiv:1007.2348 [astro-ph.CO]. 

\bibitem{Tsyba2011} P.Y. Tsyba, I.I. Kulnazarov, K.K. Yerzhanov, R. Myrzakulov, Int. J. Theor. Phys. 50 (2011) 1876, arXiv:1008.0779 [astro-ph.CO]. 

\bibitem{Dent2011} J.B. Dent, S. Dutta, E.N. Saridakis, J. Cosmol. Astropart. Phys. 1101 (2011) 009, arXiv:1010.2215 [astro-ph.CO]. 

\bibitem{Chen2011} S.-H. Chen, J.B. Dent, S. Dutta, E.N. Saridakis, Phys. Rev. D 83 (2011) 023508, arXiv:1008.1250 [astro-ph.CO]. 

\bibitem{Bengochea2011} G.R. Bengochea, Phys. Lett. B 695 (2011) 405, arXiv:1008.3188 [astro-ph.CO]. 

\bibitem{Wu2010b} P. Wu, H. Yu, Phys. Lett. B 693 (2010) 415, arXiv:1006.0674 [gr-qc]. 

\bibitem{Yang2011} R.-J. Yang, Europhys. Lett. 93 (2011) 60001, arXiv:1010.1376 [gr-qc]. 

\bibitem{Zhang2011} Y. Zhang, H. Li, Y. Gong, Z.-H. Zhu, J. Cosmol. Astropart. Phys. 07 (2011) 015, arXiv:1103.0719 [astro-ph.CO]. 

\bibitem{Li2011} B. Li, T.P. Sotiriou, J.D. Barrow, Phys. Rev. D 83 (2011) 064035, arXiv:1010.1041 [gr-qc]. 

\bibitem{Wu2011} P. Wu, H. Yu, Eur. Phys. J. C 71 (2011) 1552, arXiv:1008.3669 [gr-qc]. 

\bibitem{Bamba2011} K. Bamba, C.-Q. Geng, C.-C. Lee, L.-W. Luo, J. Cosmol. Astropart. Phys. 1101 (2011) 021, arXiv:1011.0508 [astro-ph.CO]. 

\bibitem{Boehmer2011} C.G. B{\"o}hmer, A. Mussa, N. Tamanini, Class. Quantum Grav. 28 (2011) 245020, arXiv:1107.4455 [gr-qc]. 

\bibitem{Wang2011} T. Wang, Phys. Rev. D 84(2011) 024042, arXiv:1102.4410 [gr-qc]. 

\bibitem{Daouda2011} M.H. Daouda, M.E. Rodrigues, M.J.S. Houndjo,  Eur. Phys. J. C 71 (2011) 1817, arXiv:1108.2920 [astro-ph.CO]. 

\bibitem{Nashed2013} G.G.L. Nashed, Gen. Relativ. Gravit. 45 (2013) 1887, arXiv:1502.05219 [gr-qc]. 

\bibitem{Abbas2015a}  G. Abbas, A. Kanwal, M. Zubair, Astrophys. Space Sci. 357 (2015) 109, arXiv:1501.05829 [physics.gen-ph]. 

\bibitem{KB1975} K.D. Krori, J. Barua, J. Phys. A: Math. Gen. 8 (1975) 508. 

\bibitem{Abbas2015b}  G. Abbas, S. Qaisar, A. Jawad, Astrophys. Space Sci. 359 (2015) 57, arXiv:1509.06711 [physics.gen-ph]. 

\bibitem{Batti2017} M.Z.H. Bhatti, Z. Yousaf, S. Hanif, Mod. Phys. Lett. A 32 (2017) 1750042. 

\bibitem{Mai2017} Z.-F. Mai, H. L{\"u}, Phys. Rev. D 95 (2017) 124024, arXiv:1704.05919 [hep-th]. 

\bibitem{Awad2017} A.M. Awad, S. Capozziello, G.G.L. Nashed, JHEP 07 (2017) 136, arXiv:1706.01773 [gr-qc].

\bibitem{Ganiou2017} M.G. Ganiou, C. A{\"i}namon, M.J.S. Houndjo,  J. Tossa, Eur. Phys. J. Plus, 132 (2017) 250.

\bibitem{Nashed2018} G.G.L. Nashed, S. Capozziello, Int. J. Mod. Phys. A 33 (2018) 13, arXiv:1710.06620 [gr-qc].

\bibitem{saha2018} P. Saha, U. Debnath, Advan. High Energ. Phys. 2018 (2018) 3901790, arXiv:1810.05537 [gr-qc].
 
\bibitem{saha2019} P. Saha, U. Debnath, Eur. Phys. J. C 79 (2019) 919, arXiv:1911.10908 [physics.gen-ph].

\bibitem{sasa2018} S. Iliji\'{c}, M. Sossich, Phys. Rev. D 98 (2018) 064047, arXiv:1807.03068 [gr-qc].

\bibitem{Nashed2019} G.G.L Nashed, E.N. Saridakis, Class. Quantum Grav. 36 (2019) 135005, arXiv:1811.03658 [gr-qc].

\bibitem{Awad2019} A.M. Awad, G.G.L. Nashed, W.El Hanafy, Eur. Phys. J. C 79 (2019) 668, arXiv:1903.12471 [gr-qc]. 

\bibitem{Capozziello2019} S. Capozziello, G.G.L. Nashed, Eur. Phys. J. C 79 (2019) 911, arXiv:1908.07381 [gr-qc].

\bibitem{Bernard2020} C. Bernard, M. Ghezelbash, Phys. Rev. D 101 (2020) 024020, arXiv:2001.00615v2 [gr-qc]. 

\bibitem{Singh2019} K.N. Singh, A. Banerjee, F. Rahaman, Phys. Rev. D 100 (2019) 084023, arXiv:1909.10882 [gr-qc].

\bibitem{chanda2019} A. Chanda, S. Dey, B.C. Paul, Eur. Phys. J. C 79 (2019) 502. 

\bibitem{debnath2019} U. Debnath, Eur. Phys. J. C  79 (2019) 499, arXiv:1901.04303 [gr-qc].

\bibitem{Madsen1992} M.S. Madsen, J.P. Mimoso, J.A. Butcher, G.F.R. Ellis, Phys. Rev. D 46 (1992) 1399. 

\bibitem{carr1975} B.J. Carr, Astrophys. J. 201 (1975) 1. 

\bibitem{wesson1978} P.S. Wesson, J. Math. Phys. 19 (1978) 2283. 

\bibitem{braje2002} T.M. Braje, R.W. Romani, Astrophys. J. 580 (2002) 1043, arXiv:astro-ph/0208069. 

\bibitem{linares2004} L.P. Linares, M. Malheiro, S. Ray, Int. J. Mod. Phys. D 13 (2004) 1355. 

\bibitem{Israel1966} W. Israel, Nuo. Cim. 44 (1966) 1, Erratum: 48 (1967) 463.

\bibitem{Darmois1927} G. Darmois, M{\'e}morial des sciences math{\'e}matiques XXV,
Fasticule XXV, (Gauthier-Villars, Paris, France 1927), Chap. V.

\bibitem{lanczos1924} K. Lanczos, Ann. Phys. (Berlin) 379 (1924) 518. 

\bibitem{sen1924} N. Sen, Ann. Phys. (Berlin) 378 (1924) 365. 

\bibitem{perry1992} G.P. Perry, R.B. Mann, Gen. Relativ. Gravit. 24 (1992) 305. 

\bibitem{lake1996} P. Musgrave, K. Lake, Class. Quant. Gravit. 13 (1996) 1885. 

\bibitem{rahaman2006} F. Rahaman, M. Kalam, S. Chakraborty, Gen. Relativ. Gravit. 38 (2006) 1687, arXiv:gr-qc/0607061. 

\bibitem{rahaman2009} F. Rahaman, M. Kalam, K.A. Rahman, Acta Phys. Pol. B 40 (2009) 1575, arXiv:0804.3852 [gr-qc]. 

\bibitem{usmani2010} A.A. Usmani, Z. Hasan, F. Rahaman, Sk.A. Rakib, S. Ray, P.K.F. Kuhfittig, Gen. Relativ. Gravit. 42 (2010) 2901, arXiv:1001.1415 [gr-qc]. 

\bibitem{rahaman2010} F. Rahaman, K.A. Rahman, Sk.A. Rakib, P.K.F. Kuhfittig, Int. J. Theor. Phys. 49 (2010) 2364, arXiv:0909.1071 [gr-qc]. 

\bibitem{dias2010} G.A.S. Dias, J.P.S. Lemos, Phys. Rev. D 82 (2010) 084023, arXiv:1008.3376 [gr-qc]. 

\bibitem{rahaman2011} F. Rahaman, P.K.F. Kuhfittig, M. Kalam, A.A. Usmani, S. Ray, Class. Quant. Gravit. 28 (2011) 155021, arXiv:1011.3600 [gr-qc].

\bibitem{ghosh2017} S. Ghosh, F. Rahaman, B.K. Guha, S. Ray, Phys. Lett. B 767 (2017) 380, arXiv:1511.05417 [physics.gen-ph]. 

\bibitem{Pani2009} P. Pani, E. Berti, V. Cardoso, Y. Chen, R. Norte, Phys. Rev. D 80 (2009) 124047, arXiv:0909.0287 [gr-qc].  

\bibitem{Pani2010a} P. Pani, E. Berti, V. Cardoso, Y. Chen, R. Norte, J. Phys.: Conference Series 222 (2010) 012032. 

\bibitem{Pani2010b} P. Pani, E. Berti, V. Cardoso, Y. Chen, R. Norte, Phys. Rev. D 81 (2010) 084011, arXiv:1001.3031 [gr-qc]. 

\bibitem{Abbott2016} B.P. Abbott et al., LIGO/Virgo Scientific Collaboration, Phys. Rev. Lett. 116 (2016) 061102, arXiv:1602.03837 [gr-qc]. 

\bibitem{Cardoso2016} V. Cardoso, E. Franzin, P. Pani, Phys. Rev. Lett. 116 (2016) 171101, Erratum: 117 (2016) 089902, arXiv:1602.07309 [gr-qc].

\bibitem{Chirenti2016} C. Chirenti, L. Rezzolla, Phys. Rev. D 94 (2016) 084016, arXiv:1602.08759 [gr-qc]. 

\bibitem{Bamba2013} K. Bamba, S. Capozziello, M.D. Laurentis, S. Nojiri, D. S\'{a}ez-G\'{o}mez, Phys. Lett. B 727 (2013) 194, arXiv:1309.2698 [gr-qc]. 

\bibitem{Aron2016} J. Aron, New Scientist, issue 3072, (2016) 4 May. 

\bibitem{Kubo2016} T. Kubo, N. Sakai, Phys. Rev. D 93 (2016) 084051. 

\bibitem{Ruggiero2016} M.L. Ruggiero, Int. J. Mod. Phys. D 25 (2016) 1650073, arXiv:1601.00588 [gr-qc]. 

\end{thebibliography}
\end{document}